# Searching for the Casimir Energy

Diego Pérez-Morelo[1,2,3], Alexander Stange[4], Richard W. Lally[4], Lawrence K. Barrett[4], Matthias Imboden[6], David K. Campbell[1,4,5], Vladimir A. Aksyuk[2] and David J. Bishop[1,4,5,7,8]

1. Department of ECE, Boston University, Boston, MA 02215
2. Physical Measurement Laboratory, National Institute of Standards and Technology, Gaithersburg, MD 20899
3. Institute for Research in Electronics and Applied Physics & Maryland NanoCenter, University of Maryland, College Park, MD 20742
4. Division of MSE, Boston University, Boston, MA 02215
5. Department of Physics, Boston University, Boston, MA 02215
6. EPFL, Neuchatel, Switzerland
7. Department of ME, Boston University, Boston, MA 02215
8. Department of BME, Boston University, Boston, MA 02215

**Abstract**

In this article, we present a nano-electromechanical system (NEMS) designed to detect changes in the Casimir Energy. The Casimir effect is a result of the appearance of quantum fluctuations in the electromagnetic vacuum. Previous experiments have used nano- or micro-scale parallel plate capacitors to detect the Casimir force by measuring the small attractive force these fluctuations exert between the two surfaces. In this new set of experiments, we aim to directly detect shifts in the Casimir *energy* in the vacuum due to the presence of metallic parallel plates, one of which is a superconductor. A change in the Casimir energy of this configuration is predicted to shift the superconducting transition temperature ($T_c$) because of an interaction between it and the superconducting condensation energy. The experiment we discuss consists of taking a superconducting film, carefully measuring its transition temperature, bringing a conducting plate close to the film, creating a Casimir cavity, and then measuring the transition temperature again. The expected shifts will be small, comparable to the normal shifts one sees in cycling superconducting films to cryogenic temperatures and so using a NEMS resonator and doing this *in situ* is the only practical way to obtain accurate, reproducible data. Using a thin Pb film and opposing Au surface, we observe no shift in $T_c$ greater than 12 µK down to a minimum spacing of approximately 70 nm.

**INTRODUCTION**

The Casimir force was first derived in 1948 as a result of calculating the van der Waals force using retarded potentials [1]. This is a purely quantum mechanical force that arises between two plates even when they are not electrically charged. Classically, there would be no force on the plates. However, due to quantum fluctuations and the freezing out of the long wavelength electromagnetic modes, there exists a net pressure exerting an attractive force. Experimentally, the effect has been seen using a number of microscale systems and devices [2-8]. Reference [4] discusses how the force varies with metallic conductivity of the plates, references [9] and [10] show how the effect can be used for practical applications and Kenneth, *et al.* [11] show how a repulsive force can be achieved. Additional work has also demonstrated the importance of nanopatterning [12] and magnetic effects [13].

Given that the metallic conductivity changes the magnitude of the Casimir force, the question immediately comes to mind "what happens if the plates become superconducting?" The answer sadly is "not much." The Casimir Effect averages the conductivity of the material over very large energy scales while the superconducting gap is relevant only for the far IR [14]. So, while the effect of superconductivity is very large (100 %) on the DC conductivity, it is negligible and un-measurable if averaged over the typical energy scales found in a Casimir cavity [15]. Therefore, one cannot see any effect on the measured Casimir force at transition temperature $T_c$.

However, as pointed out in work by Bimonte and co-workers [16], one might be able to see an effect of Casimir *energy* on superconductivity. In a type I superconductor, the critical parallel field $H_{c||}(T)$ is given by the change in free energy, $\Delta F$, which is the difference between the free energy in the superconducting and normal state:

$$(H_{c||}(T))^2 \propto \Delta F(T) \qquad (1)$$

Ref. [16] suggests that this condensation energy may also have a contribution due to the Casimir energy:

$$\Delta F = E_{cond}(T) + \Delta E_{Cas}(T) \qquad (2)$$

Where $E_{cond}(T)$ is due to superconductivity and $\Delta E_{Cas}(T)$ is due to the Casimir Energy. For small modulations in this Casimir term, the critical field is modulated by a factor proportional to the ratio $\Delta E_{Cas}/E_{cond}$. Calculations done in ref. [17] suggest this fraction could be as large as 10 % in certain thin film materials with low condensation energies.

The theory suggests a couple of things. The first is that placing a superconducting film in a Casimir cavity shifts $T_c$. Secondly, measuring the effect is contingent upon keeping $E_{cond}$ constant. Therefore, any attempt to compare several different films, some inside cavities and some not, may suffer from uncertainty due to variations in the highly process-dependent characteristics of superconducting thin films. Even on a single die, different regions of deposited material may display a slightly different superconducting transition temperature due to thickness variation, local roughness, or temperature gradients during deposition. In this work, we present a technique in which a NEMS structure is used to move a plate relative to a single superconducting film *in situ*. The basic concept is shown in Figure 1, in which we sit on the shoulder of the superconducting transition and actuate a nearby metallic plate, thus modulating the Casimir energy while monitoring the film resistance. Due to the sharp slope of the superconducting transition, a small change in $T_c$ (due to a change in Casimir energy) would manifest as a measureable change in sample resistance.

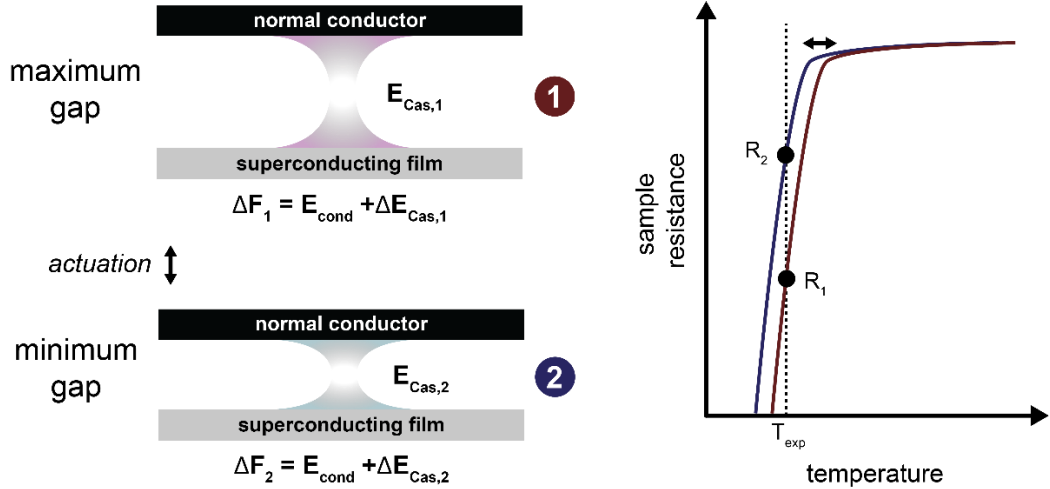

**Figure 1:** Basic concept of our experiment. Holding the system at temperature $T_{exp}$, just below the transition, a conducting plate is actuated in close proximity to a superconductor in order to modulate the Casimir contribution to its total free energy change of condensation ($\Delta F$), and thereby $T_c$. The two positions shown correspond to the maximum and minimum gap sizes due to an oscillation of the plate.

There are a number of experimental challenges posed by the concept shown in Figure 1. According to calculations done in references [16] and [17], the spacing of the Casimir cavity should be in the range of a few nanometers to 100 nm and the film thickness should be on the order of 10 nm. This places serious constraints on choice of materials, many of which tend to ball up and form islanded microstructures when thin [18]. However, evaporating onto cryogenically cooled surfaces allows for quenched condensation of the material, forming very smooth, amorphous films [19, 20]. For this reason, an *in situ* deposition method is used in which the superconducting film is deposited at the chip scale, below the superconducting transition temperature. This "fab-on-a-chip" methodology is explained further in the Methods section as well as in references [20-24]. It is with this quenched-condensed thin film (which serves as one half of a tunable Casimir cavity) that we are able to probe changes in the Casimir energy.

**RESULTS**

Experiments are performed by assembling a target chip and a source chip into a single package, cooling the system down to cryogenic temperatures, depositing a superconducting thin film onto the target die and forming a Casimir cavity, characterizing the film, then dynamically tuning the size of the Casimir cavity while monitoring the film resistance. *Ex situ* characterization of the film and cavity using scanning electron microscopy (SEM) and atomic force microscopy (AFM) is also done after the experiment is complete. The following results are presented according to this sequence.

**Chip-scale evaporation and measurement set-up**

The non-rigid Casimir cavity imposes serious nanofabrication challenges that must be overcome to successfully obtain a functional device. Firstly, due to process incompatibilities we need to be able to deposit the superconducting sample underneath the metallic plate after releasing the movable structures. Secondly, due to the oxidation of the thin film, the evaporation and resistance measurement need to be done without breaking vacuum conditions. Finally, low temperatures can be used to reduce the migration of the evaporated material along the substrate enabling high quality films.

Figure 2 shows a schematic of the three-die arrangement developed for this experiment, consisting of one NEMS target die (onto which material will be evaporated) centered between two micro-source dies. The flux of material being evaporated diagonally from each micro-source die will reach the target die and form a continuous, thin film underneath the top Au layer. This top Au layer on the target die serves as both a physical mask as well as a movable plate (in order to vary the Casimir gap size, which is nominally $g_0$). The bottom Au layer of the target die consists of two sets of electrical leads. The rectangular structures shown in figure 2 (target schematic) are the plate drive and sense electrodes for the movable gold plate, with a ground shield around them. The four leads heading off at 45° (**P1**, **P2**, **P3**, **P4**) in the target die schematic are the four-point contacts to the superconducting film used to measure its resistance. The film is formed by two angular evaporations through the holes (red) that combine to form a continuous strip along the center. Another important feature of the target die is the presence of silicon oxide pillars which serve as physical stops for the movable Au plate, both protecting the sample from contact with the Au as well as providing information about the minimum cavity size achieved. The micro-source dies, shown in purple, are separated by distance $d_{sources}$ and consist of an array of micro-scale heaters pre-loaded with a layer of superconducting material. The design and fabrication of the micro-sources and target are explained further in Supplementary Information.

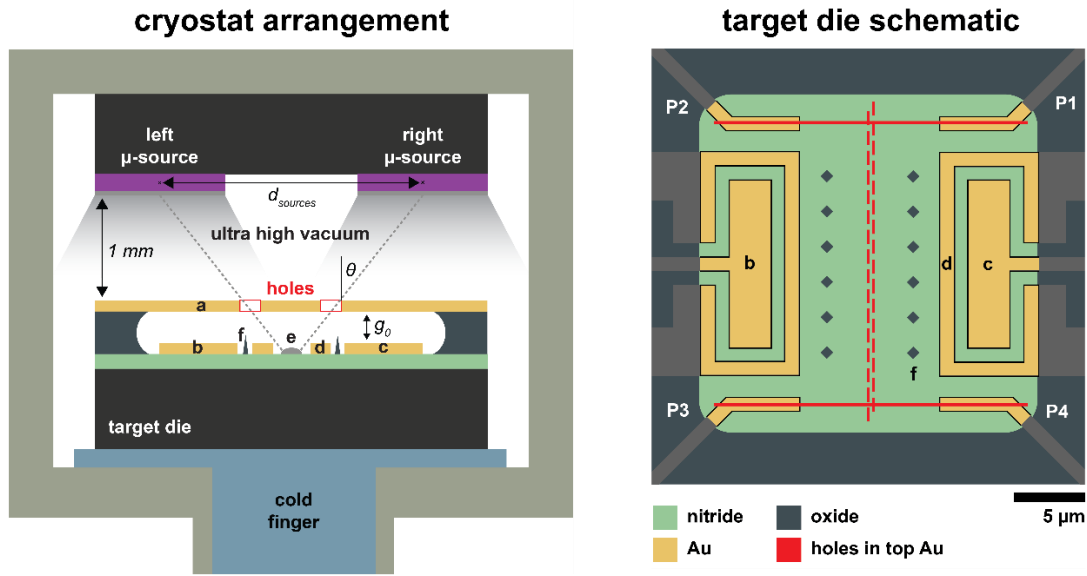

**Figure 2:** Schematic of experimental set-up and target die. Two MEMS-based micro-sources (purple) generate a flux of Pb atoms which is deposited onto a specially designed target die. Using an angled evaporation through a suspended, patterned Au layer (**a**), a continuous film (**e**) is formed which connects four pre-positioned measurement leads (**P1**, **P2**, **P3**, **P4**). Au electrodes (**b**, **c**) are then used to actuate and sense the top suspended Au layer. Grounded guards (**d**) surround each electrode to minimize current leakage and stray electric fields. As shown in the target schematic, the pattern etched into the top Au layer (red) acts as both the mask for the evaporation and allows the suspended portion of the top Au layer to move, resulting in a tunable Casimir cavity. Silicon oxide pillars (**f**) serve as a physical stop for the movable Au plate.

### Quenched-condensed Pb thin film

After the system is cooled to ≈ 3 K, the micro-sources are slowly heated until evaporation of the Pb occurs. By applying short voltage pulses through the micro-sources, very small amounts of material can be evaporated in a controlled manner. Because the target die is cooled to cryogenic temperatures, the Pb reaching the target forms a quenched-condensed film of just 20 nm to 30 nm (see *ex situ* AFM results). According to the mask pattern displayed in figure 2, the dual-angle deposited Pb film connects four Au measurement leads in an 'H' pattern, allowing for a four-point resistance measurement to be made.

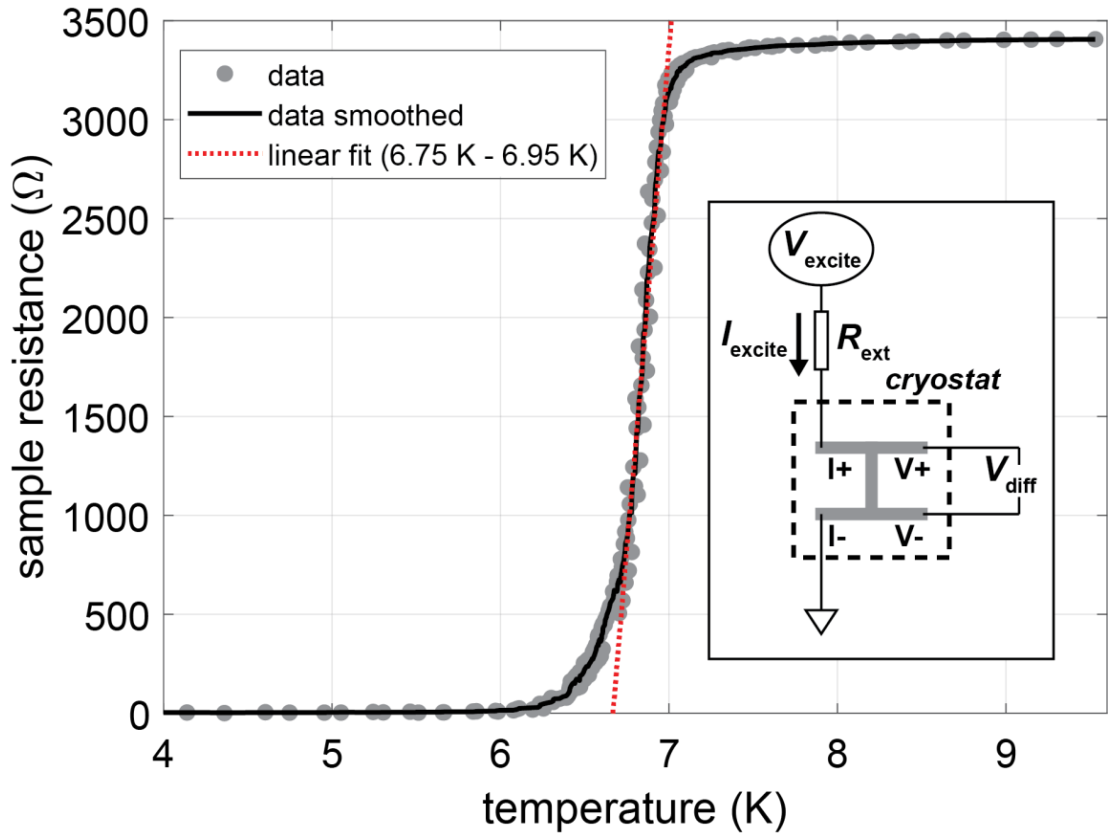

**Figure 3:** Superconducting transition of the quenched-condensed Pb film. Data points (gray dots) are taken both sweeping up and down through the transition and the black line is a smoothed average of these data points. The slope of the transition is calculated to be (11.4 ± 0.4) kΩ/K. INSET: four-point resistance detection scheme. Current is applied through an external resistance and the voltage drop across the Pb sample is measured.

Shown in figure 3 is the measured resistance of the quenched-condensed Pb film as the temperature is swept from 4 K to 9.5 K and back down. The temperature at which the film begins to condense into the superconducting phase is ≈ 7.05 K, just under the bulk value of 7.193 K [25] indicating a thicker film (around 20 nm [20]). The resistance of the film above $T_c$ is ≈ 3.4 kΩ and the slope at the center of the transition is (11.4 ± 0.4) kΩ/K. The uncertainty on the slope is reported from the 95 % confidence bounds of the linear fitting in the range of 6.75 K to 6.95 K.

### Cavity modulation and $T_c$ shift measurement

The experiment is carried out by setting the temperature of the cryostat to the steepest point of the transition (T ≈ 6.88 K), where the measured resistance is most sensitive to changes in temperature. Then, the plate drive voltage is swept through its mechanical resonance while measuring the modulation of the resistance of the sample at the mechanical motion frequency. By operating the NEMS around resonance, we are not only able to produce large changes in gap size, but also perform the measurement at a high frequency (> 1 MHz) which greatly reduces measurement noise. Plotted in figure 4 are three trials using this detection method, Trial #1, Trial #2, and Trial #3. The black data points show the frequency dependence of the amplitude and phase of the Au plate (left and right figures respectively). A common Duffing-nonlinear oscillator response is evident for amplitudes below the plate mechanical contact with the oxide pillars. The blue data points show the change in resistance of the sample expressed in units of change in the transition temperature, using the slope calculated in figure 3.

The abscissa of the plots in figure 4 is the frequency of the signal being applied at the drive electrode. Because this is a purely AC signal, the electrostatic force applied to the plate and its resulting motion is actually at twice this frequency (see Methods section for details on this detection scheme). This 2x frequency difference between the electrical drive and the expected superconductor modulation signal largely eliminates any direct electrical crosstalk. For each trial, the frequency is swept upwards and the plate amplitude/phase and sample resistance are all recorded simultaneously.

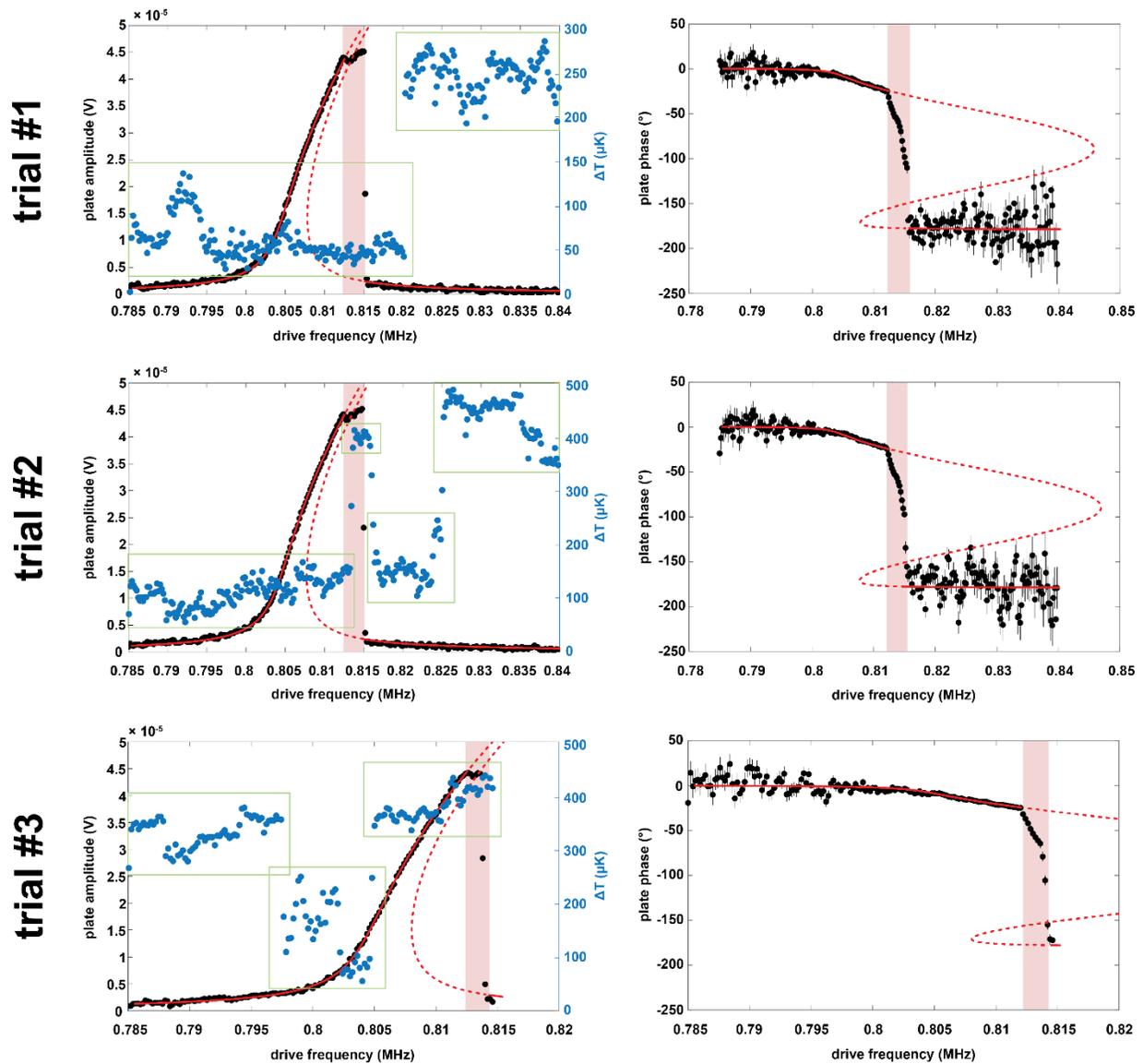

**Figure 4:** High frequency detection of superconducting Pb film and cavity size. Black data is amplitude (left plots) and phase (right plots) of the movable plate for each trial, along with fits to a Duffing model (red lines). Uncertainties for amplitude measurements are smaller than the symbol size. Phase uncertainties are one standard deviation uncertainties propagated from the measured statistical uncertainties in the x and y quadratures. Regions where the plate comes into contact with the oxide pillars are highlighted in pink. As the plate is swept through its resonance, the Pb resistance is recorded and scaled to units of temperature change (blue data) using the slope of the superconducting transition. Note that $\Delta T = 0$ µK is arbitrary from plot to plot. The green boxes indicate regions of data which are used to quantify the one standard deviation uncertainty of the $\Delta T$ measurements (28 µK for each point).

Also plotted in figure 5 are the same three trials shown in figure 4 but zoomed in to the region of interest (close to contact with the oxide pillars and just after the amplitude drop). In addition to plotting the data points, we include two sliding average curves to help visualize trends in the noisy data. These sliding averages take the mean of either one value or three values on either side of each data point (thin blue line and thick purple line respectively).

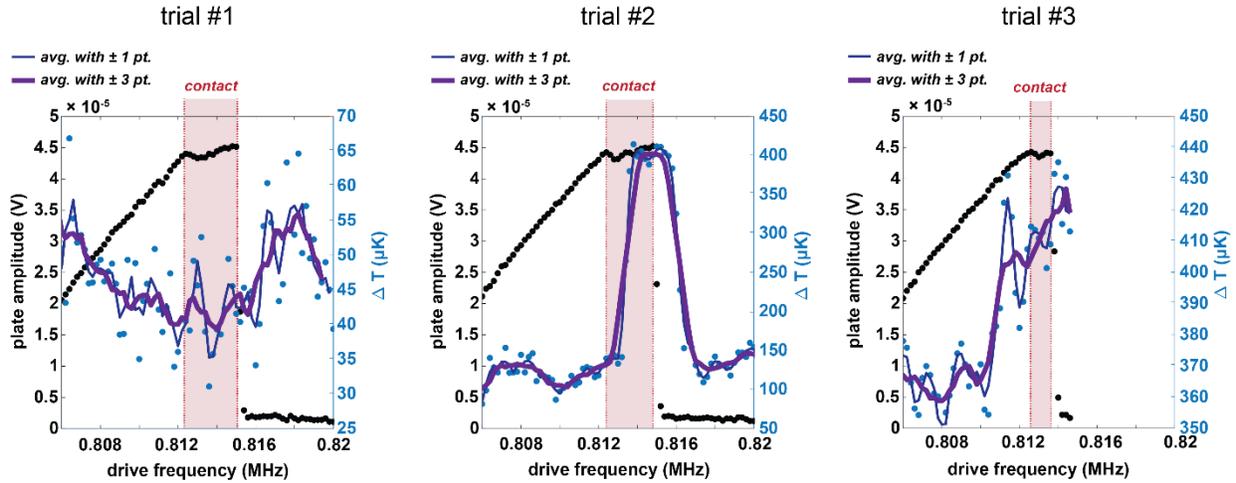

**Figure 5** A closer look at high frequency results. Only the regions before, during, and after contact are plotted. In addition to data points, sliding averages are plotted for better visualization. The one standard deviation uncertainty on ΔT measurements is 28 µK for each point.

## *Ex situ* characterization

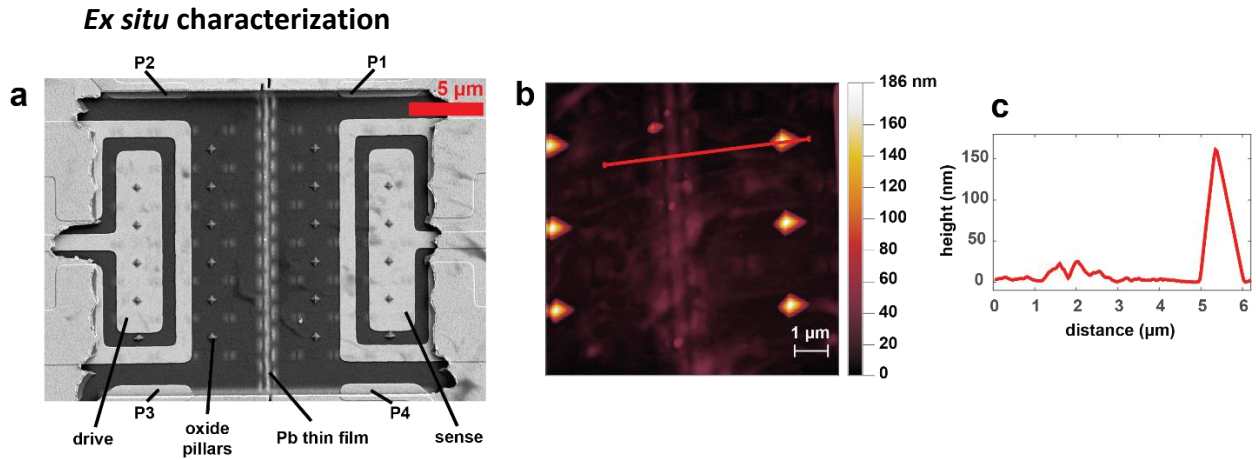

**Figure 6:** *Ex situ* analysis of the cavity and film. **a.** SEM image of cavity with the top Au layer removed using adhesive tape. A continuous Pb film can be seem connecting the four Au leads. **b.** AFM measurement of the Pb sample and several oxide stops. The height data corresponding to the red profile is plotted in **c**.

After running the experiment in the cryostat, SEM and AFM characterization of the nano-cavity (with the top Au surface removed) is performed. In Figure 6a the features of the bottom

Au layer depicted in figure 2 are shown, consisting of two rectangular electrodes and four measurement leads. Additionally, the deposited Pb sample is visible, connecting all four leads and forming a continuous strip down the center. Figure 6b and 6c show AFM height measurements of the sample and the oxide stops. The sample height is measured to vary between 20 nm and 30 nm. Using the SEM image, the length of the central portion of the sample is (20.1 ± 0.1) µm and its width is (403 ± 57) nm. These values are obtained by averaging several measurements along the sample width and length and the uncertainty is one standard deviation. The height of the oxide pillars is measured to be (160 ± 1) nm from averaging the heights of the 6 measured pillars shown in figure 6b and the uncertainty is one standard deviation.

Because the plate comes into contact with the oxide pillars, it is possible to estimate the vertical displacement of the center of the plate, however in order to do this, an assumption of the shape of the deformed Au must be made. The difficulty of this estimate is that the stress state of the suspended Au layer at cryogenic temperature is unknown. Using a series of finite element simulations, a likely range of minimum Au/Pb spacing (i.e. when the amplitude of the plate is maximum) is estimated to be between 63 nm and 73 nm. Across the entire length of the sample, the average Au/Pb spacing at maximum amplitude is estimated to lie in between 85 nm and 141 nm. This analysis is explained further in Methods.

**DISCUSSION**

We have developed a platform based on a variable gap cavity including a Au nanoscale mechanical resonator and a high-quality quenched-condensed superconductor film, enabling simultaneous electro-mechanical control of the nanometric gap between plates and transport measurement of the superconducting film using a four-probe configuration. The high frequency detection method presented makes use of the natural resonance of the plate to obtain large modulations of cavity size and allows for phase sensitive detection of changes in the sample resistance due to plate motion only.

**Analysis of results**

In this set of experiments, the temperature of the cryostat is set to the steepest part of the superconducting transition (T ≈ 6.88 K) and three long-duration sweeps are made driving the plate to resonance from low to high frequency until it abruptly loses amplitude due to its non-linear resonant response. An important feature in the plate response is the evidence in all three trials of a sudden clear deviation from the Duffing-nonlinear oscillator model in the amplitude and phase behavior occurring at around 811.2 kHz. This is due to the onset of the plate interacting with the oxide stops. Any additional non-linearities in the plate mechanics due to deformation alone would not appear in this discontinuous manner, but instead would likely appear gradually. In addition, the amplitude is nearly fixed beyond this point, which is further indication of contact. Upon contacting the pillars, the maximum peak displacement of the most deflected point on the plate center line is estimated to be between 183 nm and 193 nm, resulting in a minimum Casimir cavity size of between 63 nm and 73 nm. Along the entire length of the superconducting sample and plate center, we estimate the average gap size at closest approach to be in between 85 nm and 141 nm. After the plate reaches its maximum amplitude at resonance (around 815 kHz), the plate amplitude abruptly drops to zero and the undeflected Casimir cavity size returns to ≈ 256 nm. This jump is where we might expect to see more clearly a corresponding change in the sample resistance if there were indeed a dependence of the superconducting transition temperature on the cavity size, however no such statistically-significant correlation is observed.

Another feature of the high frequency ΔT measurements are seemingly random jumps of (300 to 400) μK. The cause of these instabilities is unknown; however, based on analyzing the three trials together, we do not believe they are related to the plate position. For example, in trial #1, we observe a large displacement at 820 kHz, which is well after the large plate oscillations have ceased. In trial #2, there is a large jump that is interestingly close to the minimum gap range, but a closer inspection of the data in figure 5 shows that it lags the plate contact by 4 or 5 data points (corresponding to 6 to 7.5 minutes of time difference), which is thus highly unlikely to be the effect we are looking for. Finally, in trial #3, ΔT appears to be tracking the plate amplitude, but then after the oscillations jump down, the signal does not follow. While undesirable, it is reasonable to assume that these intermittent jumps would not have obscured a correlation between the abrupt change in the plate amplitude and the sample measurement. Rather, it is the underlying stochastic noise, indicated by the spread of data of the individual areas shown in the green boxes in figure 4, which determines the experimental resolution.

In order to accurately quantify the uncertainty of the temperature measurement, we first consider this stochastic spread of the raw data as well as the uncertainty in the slope of the transition. For this value, we use the lower 95 % confidence bound of the slope (11 kΩ/K) in order to conservatively claim a resolution. Using one standard deviation from each dataset shown in the green areas of figure 4, averaging these with a weight prescribed by the number of data points in each set, and dividing by the lower estimate of the slope, we calculate a one standard deviation per data point of 28 μK.

We look for the change in the critical temperature at the point of the abrupt change from high amplitude vibration to almost negligible vibration of the plate. At this jump, we quantify the change in the measured data by averaging 4 individual ΔT points immediately before and immediately after the jump and subtracting the averaged values. The average difference from the three experimental runs is (7 ± 12) μK. Thus, any effect of plate position on the transition temperature of the superconductor in our system is well below the one standard deviation statistical uncertainty of 12 μK.

**Interpretation of the experiment**

In all three trials, no clear correlation between plate amplitude or position and transition temperature is observed above our measurement resolution of 28 μK, and from the three trials combined, no change in $T_c$ is observed to exceed the one standard deviation uncertainty of 12 μK. There are a few reasons which may explain this null result. No observation of this effect may be due to one or more of the following: 1). geometrical limitations, 2). superconductor limitations, 3). theoretical uncertainties.

In the first case, the device geometry may not allow for a small enough cavity to clearly observe a shift. The minimum cavity size we are able to reach with the current configuration is at best on the order of 70 nm. Although this brings us into a theoretically interesting range, it may not be small enough to produce a measurable shift with the current materials. The uncertainty in the exact gap size exists due to the uncertain stress in the structure near 7 K, which is expanded upon in the Methods section. Using a basic scaling law shown in equation 8 (derived from calculations done in ref. [17]), we estimate the relative change in Casimir free energy between this minimum cavity size and the un-deformed cavity size to be ≈ 400 %. Further optimization of the timed oxide undercut and geometric design may allow for smaller cavities, which would increase the magnitude of the Casimir free energy as compared to the condensation energy of the Pb film.

Regarding the superconducting film itself, there are certain key material properties to consider when conducting this type of experiment. First of all, a low $T_c$ value is generally desired because the condensation energy scales $\propto T_c^{2.6}$ [26]. As discussed, it is the ratio of the Casimir free energy to the condensation energy that determines the magnitude of the shift in critical field, so generally speaking, the lower $T_c$ the better. In the case of the experiment presented here, Pb has a relatively high $T_c$ value but was chosen for other experimental advantages. Future work may involve investigating lower $T_c$ materials. Another important material property is the plasma frequency. Bimonte *et al*. show that high plasma frequency materials can change the strength of the Casimir free energy term by almost one order of magnitude [17]. Many of the calculations in their work assume Be, which has a plasma frequency of around 18 eV compared to Pb which has a value of roughly 8 eV [27]. This reduction in reflectivity at higher frequencies may result in a Casimir energy contribution to the free energy of condensation that is too small to observe.

Finally, there is the question of what we expect theoretically in the limit of zero applied magnetic field. The calculation methods used in this low field limit are not possible due to the condensation energy and the change in free energy becoming comparable [28]. It is therefore not exactly clear what one might expect in terms of the magnitude of the change in critical temperature as a result of Casimir energy variation. What our experiment shows is that for our geometry, materials, and in the range of temperatures we can resolve, there is no observed effect. Most certainly, the next step in this project will be to include magnetic characterization as well. Integrating electromagnetic coils into the setup would allow for critical field shifts to be measured, and an experimental result that can be more directly compared to theory could then be obtained.

**CONCLUSIONS**

We have developed a unique nanomechanical transducer-based measurement technique and undertaken a careful series of experiments endeavoring to directly measure shifts in the Casimir energy by placing a superconducting Pb film in a cavity and tuning the gap, looking for effects on the superconducting transition temperature of the film. Our chip-scale system can deposit and measure a superconducting thin film while simultaneously actuating a nearby plate, forming a tunable Casimir cavity. The *in situ* deposition process is achieved with two arrays of MEMS heaters that have been pre-loaded with a thick film of Pb, and can be pulsed at low temperature to evaporate small amounts of material. The thin film is produced by using a shadow mask to define a precise pattern of the evaporated Pb (incident on the mask from two sides), that connects four metallic measurement leads and also creates a thin section of Pb directly underneath the movable Au plate. By driving a current with two sets of leads and measuring the voltage drop with the other two, we can measure the resistance of the Pb. We monitor this resistance as the Au plate is driven to its mechanical resonance and back to zero amplitude.

Using finite element analysis, we are able to estimate the deformed mode shape at resonance which puts the minimum separation of the Pb and Au somewhere between 63 nm and 73 nm. We estimate that at this gap size, the presence of the Au plate may be expected to produce relative changes in the Casimir free energy of the Pb film of 2x to 4x. Our results conclude that for this arrangement however, there are no observed Casimir-induced changes on the $T_c$ greater than 12 μK.

**METHODS**

### *In situ* film deposition and characterization

An experiment is performed in a closed cycle cryostat specially designed for low mechanical vibrations. A typical experiment is done the following way: 1) The source and target are mounted opposite one another and cooled to ≈ 3 K in the closed cycle system, 2) at low temperatures, Pb is evaporated onto the substrate producing a smooth, continuous superconducting film, 3) the resistance of the Pb film is measured across its transition and 4) the spacing of the Casimir cavity is modulated by applying an alternating voltage between the drive electrode and the suspended Au plate while the temperature of the system is held at the shoulder of the superconducting transition.

The resistance and transition of the Pb film is initially measured using a four-point configuration, shown schematically in figure 3 (inset). An excitation current, $I_{ex}$ is applied at lead I+ (in series with an external 1 MΩ resistor) and travels through the central portion of the sample to I-, which is grounded through a 1 kΩ resistor. A voltage difference is then measured between the other two leads, V+ and V-. $I_{ex}$ is a low frequency AC current of ≈ 5 nA at 37.7 Hz. The differential voltage measurement is measured using a lock-in amplifier and the resistance of the sample is then $V_{diff}/I_{ex}$. The temperature of the system is slowly swept across the transition (both down and up) which provides the slope of the transition, $dR_s/dT$, as well as $T_c$.

### Plate actuation and high frequency measurement

The dynamic detection scheme involves measuring $V_{diff}$ at the same frequency the plate is moving. First, we set the cryostat temperature to just below $T_c$ where the resistance measurement will be most sensitive to changes in temperature. Then, if the position of the plate does indeed influence the sample resistance due to the Casimir effect, we expect a small modulation of $R_s$ at the plate frequency. The voltage difference measured would then be:

$$V_{\text{diff}} = I_{\text{ex}} \cdot (R_{s,0} + \Delta R(t)) \qquad (3)$$

Where $R_{s,0}$ is the nominal sample resistance and $\Delta R(t)$ is the small variation due to the plate. We can further break down ΔR:

$$\Delta R = \underbrace{\frac{\partial R}{\partial T}}_{\text{transition slope}} \cdot \underbrace{\frac{\partial T}{\partial d}}_{\text{Casimir}} \cdot \underbrace{d(t)}_{\text{plate position}} \qquad (4)$$

The 'Casimir' term is the theorized change in $T_c$ of the Pb sample due to the changing position of the plate, d(t). This is what we intend to detect. Thus, if we consider only the component of $V_{diff}$ at the plate frequency, whose amplitude we will label $|V_{diff}|_{plate}$, we can re-arrange equation ΔR to obtain the amplitude of this expected effect:

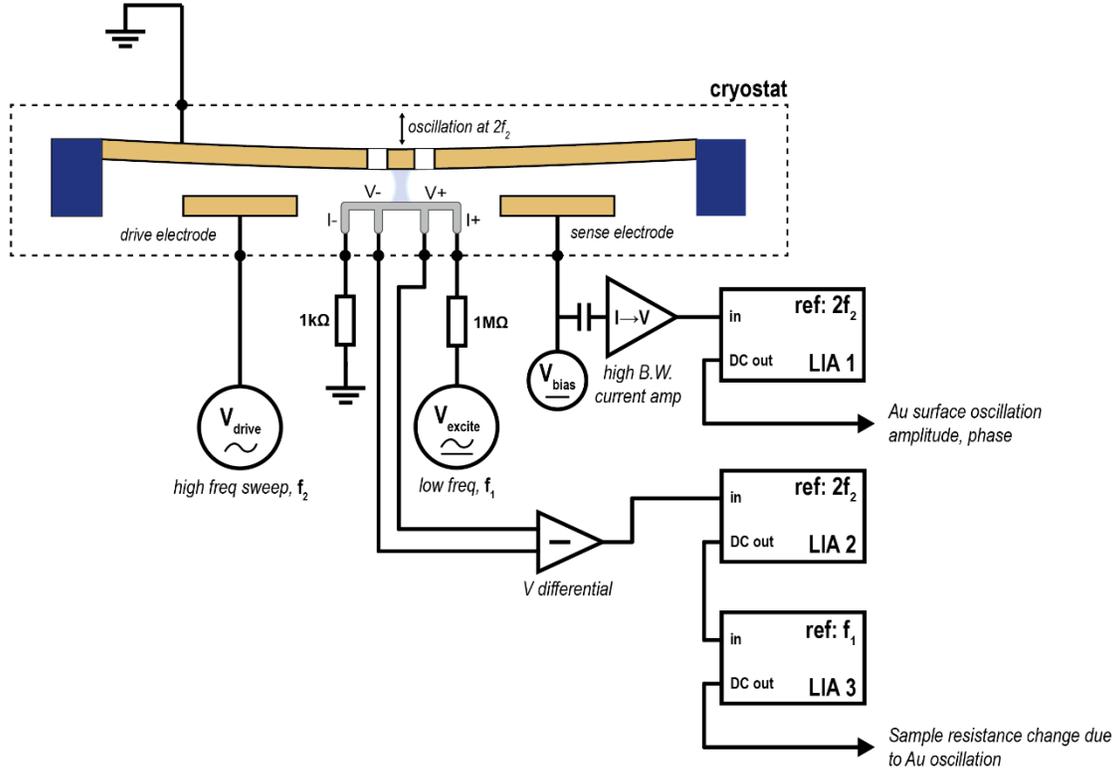

**Figure 7**: High frequency detection circuit. The cryostat is held at a constant temperature just below $T_c$ on the slope of the transition. The high-frequency $V_{drive}$ signal is applied to the drive electrode and swept at frequency $f_2$. The plate then feels an electrostatic force at frequency $2f_2$. The amplitude of the plate is monitored by measuring the AC current going through the sense electrode using a current-to-voltage amplifier and a lock-in referenced to $2f_2$ (LIA 1). The voltage drop across the Pb sample is also being detected at $2f_2$. If there is any change in $T_c$ due to the Casimir cavity size, this is the frequency it would occur at. Because the excitation current, $I_{ex}$, is alternating at $f_1$, the DC output of LIA 2 is fed into a third lock-in, LIA 3, which is referenced at $f_1$.

$$\left|\frac{\partial T}{\partial d}\right| = \frac{|V_{\text{diff}}|_{\text{plate}}}{|I_{\text{ex}}(t)| \cdot \frac{\partial R}{\partial T} \cdot |d(t)|} \qquad (5)$$

In this case, a more complex detection circuit is required. In figure 7, a high frequency AC drive signal is used to actuate the plate at its resonance. The plate amplitude is detected at twice this drive frequency using LIA 1. Simultaneously, a current is going through the sample at a low frequency, while $V_{diff}$ is being measured at the same frequency as the plate using LIA 2. The output of LIA 2 is then fed into LIA 3 which is locked into the low frequency of the excitation signal. The DC output of LIA 3 is then equal to the amplitude $|V_{diff}|_{plate}$. Table 1 reports nominal values of each parameter presented in the schematic.

| parameter name | value |
|---|---|
| $f_1$ | 37.7 Hz |
| $V_{ex}$ | 7.07 mV [DC] and 10 mV [RMS] |
| $V_{bias}$ | 4 V [DC] |
| $f_2$ range | 785 kHz to 840 kHz |
| $f_2$ frequency step | 200 Hz |
| $f_2$ time step | 90 s |
| $V_{drive}$ | 3 V [RMS] |
| LIA 1 time constant | 30 s |
| LIA 2 time constant | 10 ms |
| LIA 3 time constant | 30 s |

**Table 1:** Values used for experiments. See figure 7 for schematic.

**Variation of Casimir Free Energy due to plate motion**

It is possible to estimate the relative variation in Casimir free energy difference using the result from ref. [17]:

$$\Delta E_{Cas} \propto \frac{1}{1+\left(\frac{d}{d_0}\right)^{1.15}} \quad (6)$$

where $d$ is the separation between superconductor and metal, and $d_0$ is the nominal separation ($\approx$ 256 nm in our experiment). It should be noted however that ref. [17] approximates the relation shown in equation 6 using a system where $T_c$ = 0.5 K, the superconductor thickness is 5 nm, and $d_0$ = 8.3 nm, which is different than the system presented here. Considering the relative change in $\Delta E_{Cas}$ to be $\delta E_{Cas} = \Delta E_{Cas}(d)/\Delta E_{Cas}(d_0)$ with d(t) changing sinusoidally at ω with amplitude A (see schematic in figure 8), we find:

$$\delta E_{Cas}(t, A) = \frac{2}{1+\left(\frac{d_0+A\sin(\omega t)}{d_0}\right)^{1.15}} \quad (7)$$

To a first approximation, we assume $\delta E_{Cas}(t)$ would vary $T_c(t)$ linearly. Because the phase sensitive detection used in the experiment is only measuring the time-averaged component of the signal at ω, we can approximate equation 7 by considering only the time averaged magnitude of the first harmonic term as a function of A. Numerically solving for the magnitude of the first

Fourier coefficient of $\delta E_{Cas}(t,A)$ over the range A = 0 nm to A = 200 nm and using $d_0$ = 256 nm, we find that it can be well-approximated by a third order polynomial:

$$|\delta E_{\text{Cas}}| = c_1 A + c_2 A^3 \qquad (8)$$

where $c_1$ = 0.0225 nm$^{-1}$ and $c_2$ = 5.905 · 10$^{-9}$ nm$^{-3}$. Using equation 8, we can then estimate the relative change in Casimir free energy for a given plate amplitude. This estimate considers two parallel areas, separated by d. In reality, the interaction of the Au plate and Pb sample is not parallel, and the geometry is defined by the deformed shape of the plate. In the following section, a study of how the plate might be deforming at its resonance is discussed.

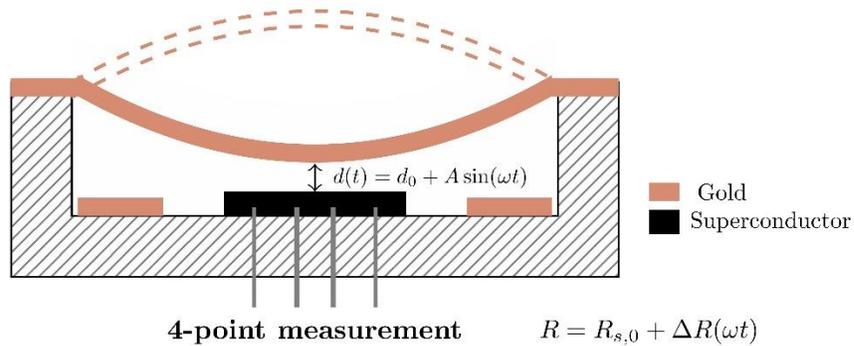

**Figure 8**: Simplified schematic of plate actuation and variable gap size.

**Estimating plate deflection and mode shape using finite element analysis**

Determining the exact amplitude of the plate and therefore the Casimir gap size is challenging, due to the significant change in dynamics that occurs between room temperature and cryogenic temperature. This is evidenced by a large change in resonant frequency, from ≈ 700 kHz to ≈ 1.8 MHz between 300 K and 3 K respectively, indicating substantial tensile stress occurring in the top Au layer due to thermal contraction of the Au relative to the substrate. By examining the behavior of the signal at the sense electrode, it is possible to determine when contact with the oxide pillars occurs, however determining the distance between the Pb sample and the center of the plate (parallel to the Pb sample) requires knowledge of the deformed shape of the Au plate at its resonance.

Using a commercial finite element software package to calculate mechanical eigenmodes, three different initial stress cases are considered in order to place bounds on the distance which the center of the plate deflects at resonance. The first case considers zero initial stress, and the second and third case include two different values of a uniaxially applied initial stress (0.1 GPa and 0.26 GPa).

After a mode shape is obtained, it is scaled in amplitude (Z) until any part of the surface comes into contact with any of the oxide pillars, and then the profile parallel to the sample is extracted. Values of maximum separation, minimum separation, and average separation are reported as well as two lengths: $L_{10}$ and $L_{25}$, which are the lengths of the sample which are within 10 % and 25 % respectively of the minimum gap size.

**Case 1:** Although there is clearly evidence of significant initial stress in the movable Au layer, it is necessary to estimate an upper bound of plate deflection by considering the zero stress case. Solving for the first eigenfrequency of the device geometry returns a value of 470 kHz, which is, as expected, well below the measured 1.8 MHz.

**Case 2:** Clearly, in order for the simulation to match the measured resonant frequency, a pre-stress must be applied. The reason for the two different stress cases 2 and 3 is that in the high stress limit, finite element analysis suggests that the fundamental mode and the second mode become near degenerate and mix due to the slight layout asymmetry in the center cut of the moving plate. This resulting high stress mode shape is essentially only one side of the cavity moving up and down, while the other side moves with much smaller amplitude. Case 2 considers an intermediate stress case before this degeneracy is reached, and the shape of the fundamental mode still resembles that of the zero stress case. In this simulation, the uniaxial pre-stress is equal to 0.1 GPa and the resulting resonant frequency is 1.12 MHz.

**Case 3:** In this simulation, the uniaxial pre-stress is increased until we reach a resonant frequency that matches experiment (≈ 1.8 MHz). This stress is equal to 0.26 GPa and is generally consistent with the estimates from the differential thermal expansion. However, in this high stress condition, the mode shape is very asymmetric due to the mixing of the fundamental and the second modes. As a result, only one edge of the Au gets close to the sample (see figure 9).

|  | min. gap | max. gap | avg. gap | $L_{10}$ | $L_{25}$ |
| --- | --- | --- | --- | --- | --- |
| **Case 1** | 63 nm | 93 nm | 85 nm | 1.2 µm | 4.6 µm |
| **Case 2** | 73 nm | 131 nm | 118 nm | 0.8 µm | 2.0 µm |
| **Case 3** | 66 nm | 169 nm | 141 nm | 0.5 µm | 1.3 µm |

**Table 2**: Results from finite element analysis for gap sizes achieved along the length of the Pb sample (20 µm in length) upon contact between the deformed Au plate and the oxide pillars. With no actuation, the gap size is assumed to be 256 nm (sample thickness subtracted from oxide thickness).

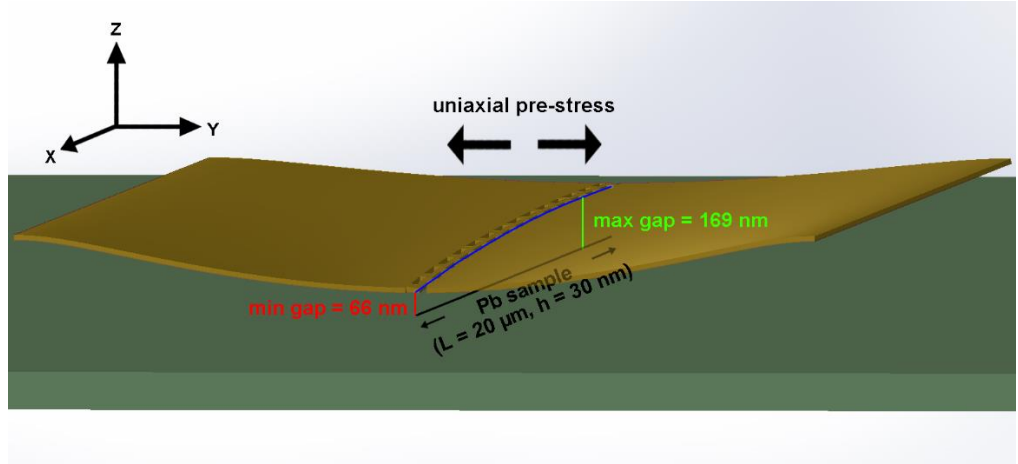

**Figure 9**: 3D representation of gap size estimation using finite element results for case 3 (high uniaxial stress). In this case, the profile of the plate along the sample (blue line) is asymmetric and reaches much closer to the sample on one side compared to the other. This reduces the average gap size as well as $L_{10}$ and $L_{25}$.

A full understanding of the exact shape of the Au plate at its resonance is not possible; however, by using the measured current amplitude from the experiment in conjunction with post-experimental finite element analysis, we are able to infer that the minimum separation of the Pb sample and the Au plate is likely in between 63 nm and 73 nm and the average separation across the length of the sample is likely in between 85 nm and 141 nm. Using the estimated scaling dependence shown in equation 8, these deflections would produce relative changes in the Casimir free energy in the Pb sample between 415 % and 439 % (when considering only the area near the minimum gap) and between 260 % and 388 % when considering the average gap over the length of the sample.

# Supplementary information

**Target die design and fabrication**

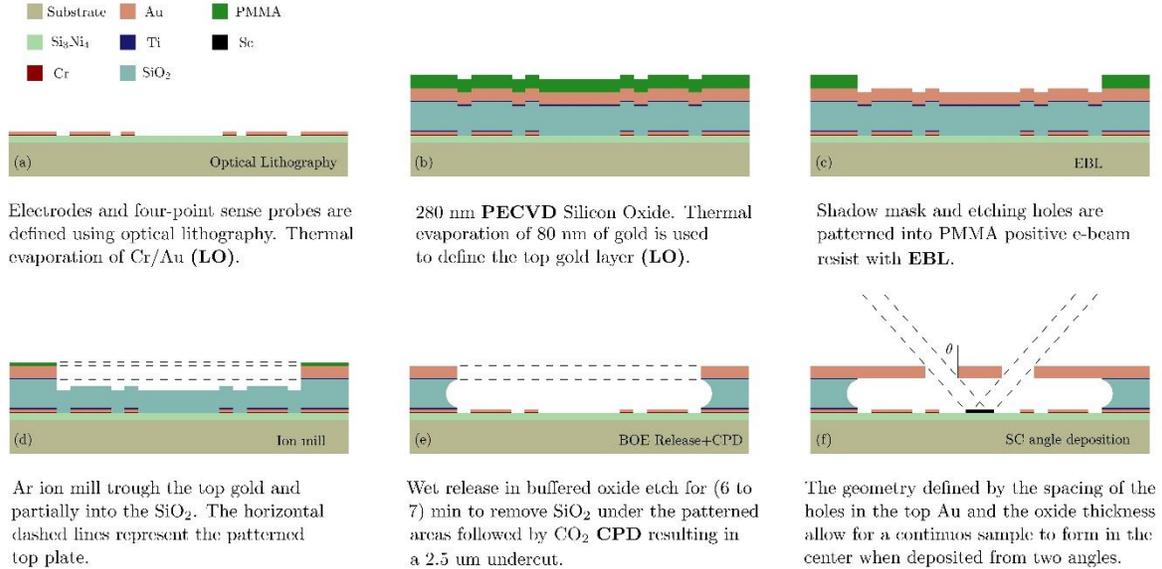

**Figure S1**: Fabrication steps of the nano-mechanical target die.

The target die contains the substrate with pre-positioned electrical leads along with a movable Au plate suspended above. This Au plate serves as both the *in situ* evaporation mask as well as tunable half of the Casimir cavity. Detailed fabrication steps can be found below as well as in ref. [1].

Figures S1a-e illustrate the process flow that has been used in the nanofabrication of the device, which leverages non-contact optical and electron-beam lithography steps to define the devices features. Devices are fabricated monolithically from bare Silicon wafer. In the first step, we deposit nominally 100 nm of Silicon Nitride using low pressure chemical deposition (LPCVD) as electrical isolation between the metallic layers and the substrate. Next, the bottom metal layer consisting of nominally 10 nm Cr, 40 nm Au, and 2 nm Ti is defined using optical lithography followed by electron beam evaporation and lift-off (LO). This layer defines two electrodes for electrostatic actuation and four-probe connections for *in situ* monitoring of the superconductor electrical resistance. Electrical leads wire each feature to bonding pads in a way that each element can be accessed independently.

Next, plasma-enhanced chemical vapor deposition (PECVD) at 180 °C is used to deposit $SiO_2$. The oxide was used both as a sacrificial layer and structural layer to define the oxide pillars.

Following the oxide deposition, a second metallic layer consisting of nominally 2 nm Ti and 80 nm Au is formed in the same way as the first layer. Au windows positioned above electrical pads are obtained from by this lift-off step, which are later used as a shadow mask to prevent shorting during the superconductor evaporation.

The top Au is then lithographically patterned for the second time using poly (methyl methacrylate) (PMMA) positive electron beam resist and electron beam lithography (EBL). The pattern is then transferred through to the Au layer with anisotropic argon ion milling, with the wafer rotating, the stage cooled to 10 °C and the incidence angle 10° off normal [1].

Finally, 1.25 mm x 1.25 mm chips, each containing one centered device, are singulated using a dicing saw. Movable structures in the top Au layer were released by wet etching in a buffered oxide etch (BOE) 6:1 solution for (6 to 7) min followed by sequential rinsing in water and isopropyl alcohol (IPA) baths. The resultant Silicon Oxide lateral undercut distance is about 2.5 µm. Finally, to prevent the suspended top gold from sticking to the bottom gold, the device was dried in a $CO_2$ critical point dryer (CPD). Note that after BOE step both Ti layers are wet etched from bottom and top of the corresponding Au layers.

In figure S1f, the final target structure and corresponding geometry is presented. The EBL pattern etched into the top Au plate (shown in black) is designed to form a continuous thin film of Pb in a 'H' shaped pattern (see figure 2 in the main text) for an angled deposition at $\theta = 35°$. This corresponds to a center-to-center spacing of the micro-source of 1.1 mm. The 'H' pattern will connect the four-point measurement leads as well as create one continuous line of Pb down the center of the cavity, beneath a continuous section of suspended Au.

**Micro-source fabrication, preparation, and configuration**

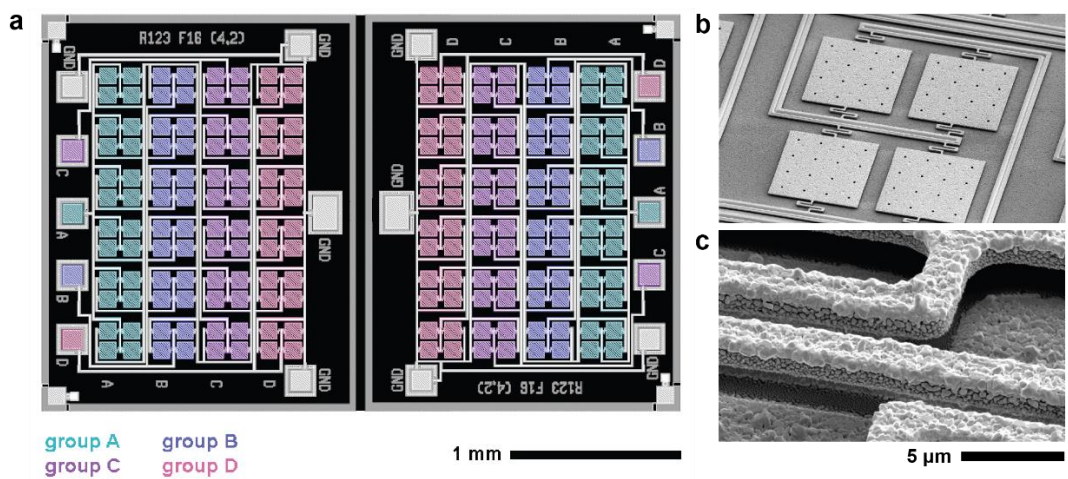

**Figure S2:** Micro-sources for quenched-condensed Pb deposition. **a.** Images of MEMS design files arranged to depict the actual micro-source configuration in the experiment. Two sets of dies, a left and a right, each contain 4 individually addressable groups of heaters, labelled A, B, C, D and color coded. By pulsing current through each group to ground (white bonding pads), we can sequentially deposit Pb from large angles towards small angles. **b.** SEM image of one set of 4 heaters. Current runs from the center of the quadrant in parallel through each heater to ground through serpentine connections. **c.** Zoomed in SEM image of the serpentine connection with Pb already loaded. The light material laying on top is Pb (about 750 nm thick) and the suspended poly-silicon structure can be seen beneath it.

The evaporation of the Pb is done from two different dies, a left and a right. By adjusting the center-to-center spacing between the left and right die, one can tune the angle of evaporation through the mask (see figure 2). The mask geometry used on the target die is designed for a center to center spacing of ≈ 1.1 mm. Each source die is 2.5mm × 2mm and consists of four individually addressable groups, each containing 24 micro-source heater units. Current is run in parallel through each group of 24 heaters individually. Figure S2a shows a colorized mask design file depicting the layout of the two micro-source dies and each of the 8 groups. Fabrication of these micro-structures is detailed in ref. [2].

Each micro-source unit consists of a freely suspended square MEMS plate, 100x100 µm$^2$ in area and 1.5 µm thick, connected with serpentine electrical leads (see figure S2b). The plates are only weakly thermally connected to the substrate they sit on and small amounts of power, typically using pulse width modulation (PWM), can be used to evaporate a small number of atoms per pulse. This can be done at low temperatures, with good control and minimal heating of the substrate upon which they land.

The actual center-to-center spacing of the micro-source dies is around 1.9 mm and the arrays are spaced over a distance of 2 mm which results in an angular deposition range of 24° to 58°. The flux reaching the target at around θ = 35° is what contributes to the Pb side of the cavity, while the flux at smaller and larger angles contributes to making a good connection between the Pb film and the four Au measurement leads. Prior to loading into the cryostat, 500 nm to 1000 nm of Pb is evaporated onto the micro-source dies by heating a resistive thermal crucible inside a vacuum chamber.

The technique used to evaporate Pb from the micro-sources involves applying very small, very short pulses of power to the heaters in order to sublime very small volumes of material from the evaporated Pb. In this regime, very low atomic fluxes can be achieved, and high quality, thin films can be deposited on the target. Voltage pulses are applied to each group of sources sequentially, starting from the outside (groups A) and moving inwards towards groups D (see figure S2a).

**References for supplementary information:**